\documentclass{elsart}
\begin{document}
\begin{flushright}
SNUTP 98-062\\
hep-ph/9806448\\
June 1998
\end{flushright}
\begin{frontmatter}
\title{Nonperturbative $O(m_c^{-2})$ Effects in $B\rightarrow X_s
\gamma$
\\ 
from Heavy Quark Effective Theory}
\author{Junegone Chay\thanksref{EMAIL}}
\thanks[EMAIL]{e-mail address: \tt chay@kupt.korea.ac.kr}
\address{Department of Physics,  Korea University, Seoul 136-701, 
Korea}
\begin{abstract}
The nonperturbative contribution, suppressed by powers of the charm
quark mass $m_c$, in the inclusive decay $B\rightarrow X_s \gamma$ is
analyzed in the context of heavy quark effective field
theory (HQEFT). According to previous analyses, the leading effects
of $O(1/m_c^2)$ arise from an external gluon attached to a charm quark
loop and can be expressed in terms of the chromomagnetic
interaction of the $b$ quark inside the $B$ meson. 
This is also true at leading order in the
HQEFT approach. However the structure of higher-dimensional operators
is different because the effects of external gluons alone do not give
a complete set of operators. A systematic method to derive
all the high-dimensional operators can be  obtained in the HQEFT
scheme using the operator product expansion. 
\end{abstract}
\end{frontmatter}

The inclusive radiative decay $B\rightarrow X_s \gamma$ has been
considerably investigated experimentally and theoretically. The CLEO
collaboration \cite{cleo} reported the branching ratio for this decay
to be ${\rm B} (B\rightarrow X_s \gamma) = (2.32 \pm 0.57 \pm 0.35)
\times 10^{-4}$. Theoretically this type of rare decay modes can
be a sensitive probe for new physics beyond the standard model. And
the photon energy spectrum is also interesting in understanding CP
violation and new physics in $B$ decays \cite{kagan}. 
In order to probe new physics, there should be a precise theoretical 
calculation of the branching ratio in the standard model. For the
decay $B\rightarrow X_s \gamma$, the full next-to-leading order
calculation in the standard model is completed 
combining matching conditions \cite{matching}, matrix
elements \cite{mat} and anomalous dimensions \cite{anom} in the
effective Hamiltonian approach.

Since the advent of the heavy quark effective field theory (HQEFT),
the understanding of the decay of $B$ mesons has come to a more
advanced level. Applying the HQEFT to  $B$ mesons, the inclusive decay
rates such as  $B \rightarrow X_c \ell \overline{\nu}$ and $B
\rightarrow X_s \gamma$ can be obtained by 
simply calculating the corresponding quark decay rates at leading
order \cite{chay}. The corrections to the leading result can be
systematically incorporated in the scheme of the HQEFT. These
corrections consist of a double series expansion both in the strong
coupling constant $\alpha_s$ and in the inverse power of the $b$ quark
mass, $m_b$. Systematic analyses of the nonperturbative effects
suppressed by powers of $m_b$ have been considered extensively in
various inclusive $B$ decays \cite{bnon}. With all these ingredients
the theoretical uncertainty is reduced to a 10\% level. The object of
this letter is to elaborate on the contribution of high-dimensional
operators, especially of order $O(1/m_c^2)$  in the context of the
HQEFT.

The theoretical framework to analyze the inclusive decay $B
\rightarrow X_s \gamma$ is to use the effective weak Hamiltonian
\cite{grin} which is given by
\begin{equation}
H_{\mathrm{eff}} = \frac{4G_F}{\sqrt{2}} V_{cs}^* V_{cb} \sum_{i=1}^8
C_i (\mu) O_i (\mu),
\label{hamil}
\end{equation}
where $O_i$ are various operators. The operator $O_2$ is the
four-quark operator given by $O_2
= (\overline{s}_{L\alpha} \gamma_{\mu}
b_{L\beta})(\overline{c}_{L\beta} \gamma^{\mu} c_{L\alpha})$ in which
$\alpha$, $\beta$ denote color indices. The operator $O_1$ differs
from $O_2$ only in the way the color indices are contracted. Operators
$O_3$ to $O_6$ are four-quark operators including all flavors below
the renormalization scale  $\mu$ of order $m_b$. The operator $O_7$ is
given as $O_7 = (e/16\pi^2) m_b \overline{s}_{L} \sigma^{\mu\nu}
F_{\mu \nu} b_R$, and $O_8$ is obtained from $O_7$ by replacing
$eF_{\mu\nu}$ by $gG_{\mu\nu}$.  

The contribution of these operators to the inclusive $B
\rightarrow X_s \gamma$ decay rate is systematically calculable.
In the leading log approximation, the matrix element of $C_7 (\mu)
O_7(\mu)$ dominates in evaluating the decay rate
$B\rightarrow X_s \gamma$ for large enough photon energies. At
next-to-leading order, other operators can contribute. In analyzing
the contribution of other operators, when it is
possible to use the operator product expansion (OPE), higher-order
processes can be described by products of some coefficients which are 
calculable in perturbation theory, and local
operators whose matrix elements represent long-distance effects.  

When operators in $H_{\mathrm{eff}}$ other than $O_7$ are included, if
we try to calculate perturbatively the decay rate for $B \rightarrow
X_s \gamma$, many problems arise because the decay rate receives
contribution from those in which the photon couples to light quarks. 
Actually the amplitude for $b\rightarrow sg\gamma$ with a light quark
loop is nonlocal. Therefore for light quarks coupled to a photon, the
idea of the OPE is not applicable.
That is, there is no OPE to parameterize nonperturbative effects from
the photon coupling to light quarks in terms of $B$ meson matrix
elements of local operators. 

However there are nonperturbative effects arising  
from the photon coupling to the charm quark, and it is possible to
have $B$ meson matrix elements of local 
operators which are suppressed by $(\Lambda_{\mathrm{QCD}}/m_c)^2$
rather than $(\Lambda_{\mathrm{QCD}}/m_b)^2$. We can 
calculate the nonperturbative effects suppressed by $1/m_c^2$ because 
the charm quark is heavy ($m_c \gg \Lambda_{\mathrm{QCD}}$). It is
exactly the reason why we can employ the HQEFT idea in order to obtain
high-dimensional operators systematically.

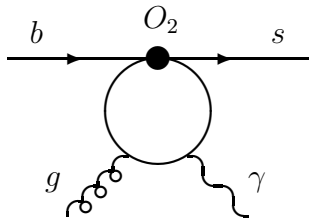
\begin{figure}[h]
\unitlength 0.1cm
  \begin{center}
    \begin{picture}(50,40)(10,0)
\thicklines
\put(20,27){\line(1,0){40}}
\put(20,27){\vector(1,0){10}}
\put(23,29){$b$}
\put(40,27){\vector(1,0){10}}
\put(55,29){$s$}
\put(40,20){\circle{14}}
\put(25,10){$g$}
\put(52,10){$\gamma$}
\put(40,27){\circle*{3}}
\put(38,31){$O_2$}
\multiput(44,12)(4,-4){2}{\oval(4,4)[tr]}
\multiput(48,12)(4,-4){2}{\oval(4,4)[bl]}
\multiput(36,12)(-2,-2){4}{\oval(4,4)[tl]}
\multiput(34.4,11.3)(-2,-2){3}{\circle{1.4}}
    \end{picture}
\end{center}
\caption{Feynman diagram for $b\rightarrow sg\gamma$ from $O_2$ in
which a photon and a gluon are attached to the $c$ quark loop. The
diagram with a photon and a gluon switched is omitted.}
\label{extglu}
\end{figure}

Naively we expect that the high-dimensional
operators, which result in nonperturbative effects, are suppressed by 
powers of $m_b$ since the only physically relevant scale in the
inclusive decay $B \rightarrow X_s \gamma$ is the mass of the $b$
quark. However Voloshin \cite{voloshin} has observed
that there are also nonperturbative effects which are suppressed by 
powers of $m_c$, the charm quark mass. This
nonperturbative effect arises from the operator $O_2$ as shown in
Fig.~\ref{extglu}, which is produced by
the quantum fluctuation of the $c\overline{c}$ pair emitting a hard
photon, and an external soft gluon is attached to the charm quark
loop. Here the blob represents the operator $O_2$.

If it is possible to use the OPE in obtaining high-dimensional
operators from the effective Hamiltonian in Eq.~(\ref{hamil}), the 
contribution from the operator $O_2$ is the most important one because 
the coefficient $C_2$ is large compared to $C_3$--$C_6$. The
high-dimensional operators arising from the operator $O_1$ starts from
the operator with more than one gluon due to the color structure,
hence more suppressed compared to the operators obtained from $O_2$.

To first order in the gluon momentum, the new effective Hamiltonian 
for $b\rightarrow s\gamma g$ is written as
\begin{equation}
H (b\rightarrow s\gamma g) = \frac{egQ_c}{48\pi^2 m_c^2}
\frac{G_F}{\sqrt{2}} C_2
V_{cs}^* V_{cb} \Bigl(\overline{s} G_{\alpha \beta} \gamma_{\mu}
(1-\gamma_5) b \Bigr) 
\epsilon^{\lambda \mu \alpha \sigma}D^{\beta}F_{\sigma \lambda}.
\label{new}
\end{equation}
Here $G_{\alpha \beta} = G_{\alpha \beta}^a T_a$ is the gluon field
strength tensor, $F_{\sigma \lambda}$ is the electromagnetic field
strength tensor, and $D^{\alpha} = \partial^{\alpha} + ieQ_c
A^{\alpha} +igG^{\alpha}$ is the covariant derivative. $Q_c=2/3$ is
the electric charge of the charm quark. And the sign
convention is $\epsilon^{0123}= +1$. 
There has been much effort to estimate the size of this new type of
nonperturbative effects in various $B$ decays such as $B\rightarrow
X_s \gamma$, $B\rightarrow X_s \ell^+
\ell^-$ \cite{ligeti,grant,wyler,rey}. 

If we keep more powers of external gluon momentum, or if we attach 
more gluons to the charm quark loop, we can obtain higher-dimensional
operators. These higher-dimensional operators are suppressed by more
powers of $m_c$ compared to the leading operator given in
Eq.~(\ref{new}). These higher-order effects are quite small compared
to the contribution of the leading operator, not because the
expansion parameter $m_b \Lambda_{\mathrm{QCD}}/m_c^2$ is small, but
because the numerical coefficients of these operators are small
\cite{ligeti,grant}.

Now we consider how we can implement the idea of the HQEFT and the OPE
to derive high-dimensional operators suppressed by powers of $m_c$ in
a systematic way. In the heavy quark limit $m_b \rightarrow \infty$,
the $b$ quark interacts with the light 
degrees of freedom, which is characterized by the residual momentum
$k$ of the $b$ quark ($p_b^{\mu} = m_b v^{\mu} +k^{\mu}$). As the
operator $O_2$ acts, the $b$ quark turns into an $s$ 
quark and a $c\overline{c}$ pair. When the virtual  $c\overline{c}$
pair emits a hard photon and soft gluons, this process induces
high-dimensional operators. Now let us also take the heavy quark limit
$m_c\rightarrow \infty$ while $m_c/m_b$ held fixed. Then when the $b$
quark turns into a $c\overline{c}$ pair, the $c\overline{c}$ pair is
still immersed in the same light degrees of freedom from the $b$
quark. Therefore the $c$ quark also carries the residual momentum $k$
of the original $b$ quark. The fact that the charm quark should have a
residual momentum is manifest if we consider the corresponding Feynman
diagram in the full theory with $W$ exchange.

From this process mentioned above, we can systematically obtain all
the high-dimensional operators in powers of $1/m_c$ by
expanding the residual momentum in the $c$ quark and using 
the OPE. There are also other contributions such as the effect of the
residual momentum in the $s$ quark, but the operators obtained from
the residual momentum of the $s$ quark are suppressed by powers of
$m_b$. Since we are interested in the operators suppressed by powers
of $m_c$ here, we will neglect all the operators suppressed by
$m_b$. And at this leading order, the $b$ field can be regarded as a
heavy field $b_v$ with momentum $p^{\mu}_b = m_b v^{\mu} +k^{\mu}$ in
the HQEFT.

These high-dimensional operators include the operators 
obtained in Ref.~\cite{voloshin,ligeti,grant,wyler,rey}, which
correspond to the so-called ``one-gluon matrix element''
\cite{bnon}. The operators from one-gluon matrix element are those in
which an external gluon is attached to the charm quark loop. In order
to make the resultant operator gauge invariant, we combine one gluon
momentum operator and the external gluon field to form the gluon field 
strength tensor $G_{\alpha \beta}$. Therefore these operators are
proportional to $G_{\alpha \beta}$, and its derivatives if there are
additional gluon momenta at higher orders. If we attach more external
gluons, we obtain high-dimensional operators in powers of $G_{\alpha
\beta}$. However, in the HQEFT, there are also other operators
which are not proportional to $G_{\alpha \beta}$ and its derivatives,
and these operators can be obtained systematically using the HQEFT and
the OPE. 
Interestingly enough, the leading contribution in the HQEFT is the
same as the operator from the one-gluon matrix element. Therefore the
numerical analysis for the contribution of the leading operator at
$O(1/m_c^2)$ in Ref.~\cite{voloshin,ligeti,grant} does not change. The
contribution of the leading operator in Eq.~(\ref{new}) compared to
the contribution of $O_7$ alone is about $\delta \Gamma /\Gamma
\approx 3$\% \cite{rey}.  

\begin{figure}[h]
\unitlength 0.1cm
  \begin{center}
    \begin{picture}(50,40)(10,0)
\thicklines
\put(20,27){\line(1,0){40}}
\put(20,27){\vector(1,0){10}}
\put(23,29){$b$}
\put(11,23){$m_b v+k$}
\put(40,27){\vector(1,0){10}}
\put(55,29){$s$}
\put(55,23){$m_b v -q +k$}
\put(40,20){\circle{14}}
\multiput(40,12)(0,-4){2}{\oval(1.5,2)[l]}
\multiput(40,10)(0,-4){2}{\oval(1.5,2)[r]}
\put(37,2){$q,\lambda$}
\put(42,9){$\gamma$}
\put(22,15){$l+k$}
\put(49,15){$l-q+k$}
\put(40,27){\circle*{3}}
\put(38,31){$O_2$}
    \end{picture}
\end{center}
\caption{Feynman diagram for $b\rightarrow s\gamma$ to obtain
high-dimensional operators from the expansion of the residual momentum
in the charm quark loop using the OPE.} 
\label{otwo}
\end{figure}
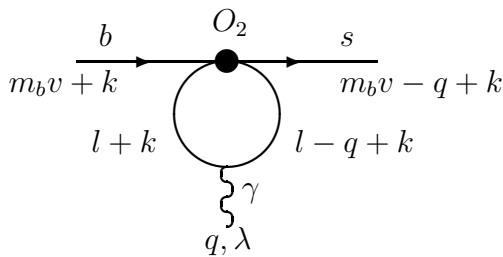

In order to systematically calculate high-dimensional operators
suppressed by powers of $m_c$ in the HQEFT, consider the Feynman
diagram shown in Fig.~\ref{otwo}. Though the operator $O_2$ is shown
in Fig.~\ref{otwo}, note that the operator $O_1$ 
should be included also when there are two or more soft gluons are
emitted. However the contribution of $O_1$ appears at next-leading
order compared to that from $O_2$. The integral in evaluating the
Feynman diagram in Fig.~\ref{otwo} can be written as   
\begin{equation}
I^{\lambda \mu} = -ieQ_c \int \frac{d^4 l}{(2\pi)^4} \mathrm{tr} \Bigl[
\frac{1}{\FMslash{l} -\FMslash{q} +\FMslash{k} -m_c} \gamma^{\lambda}
\frac{1}{\FMslash{l} +\FMslash{k} -m_c} \gamma^{\mu} (1-\gamma_5)
\Bigr],
\label{sloop}
\end{equation}
where $q$ denotes the photon momentum.

Now we expand  $I^{\lambda \mu}$ in powers of the residual momentum
$k$ and collect the terms which are suppressed by powers of $m_c$. We
use the operator product expansion in order to obtain the 
corresponding operators. That is, we replace
$k^{\mu}$ by $\Pi^{\mu} = iD^{\mu}$, where  $D^{\mu}$ 
is the covariant derivative acting on the $b$ field. The operator
$\Pi^{\mu}$ satisfies the commutation relation $[\Pi^{\mu},
\Pi^{\nu}]= -ig G^{\mu \nu}$. The amplitude $M^{\lambda}$ for
$b\rightarrow s\gamma$ + many gluons from Fig.~\ref{otwo} can be
written in terms of $I^{\lambda \mu}$ as 
\begin{equation}
iM^{\lambda} = \frac{G_F}{\sqrt{2}} V_{cs}^* V_{cb} C_2 \Bigl(
\overline{s} I^{\lambda \mu}  \gamma_{\mu} (1-\gamma_5) b\Bigr).
\end{equation}

At a first glance, the integral in
Eq.~(\ref{sloop}) seems to be made independent of the residual
momentum when we shift the loop momentum variable $l$ by $l+k$, hence 
appear no high-dimensional operators. But this
shift of the integration variable is not allowed in the integral
because the integral as such is quadratically divergent. However, if
we employ the dimensional regularization to regulate the integral,
only the $\gamma_5$-dependent part of the integral should be treated
carefully. 

The $\gamma_5$-independent integral can be shifted to be
independent of the residual momentum, hence producing no
high-dimensional operators. Furthermore after the shift of the
integration variable, the $\gamma_5$-independent integral vanishes for
the on-shell photon. Therefore to all orders there are no
high-dimensional operators from the $\gamma_5$-independent part. It
means that high-dimensional operators suppressed by powers of $m_c$ in
$B\rightarrow X_s \gamma$ should always include the
Levi-Civita $\epsilon$ tensor which
appears by taking the trace of Dirac matrices involving $\gamma_5$ in
Eq.~(\ref{sloop}).  

In order to calculate the remaining $\gamma_5$-dependent part which is
suppressed by powers of $m_c$, we
first expand the integrand in powers of $k$, and we can shift the
integration variable only when the integral is guaranteed to be
convergent. In the resultant expression, we get high-dimensional
operators by replacing $k$ with the covariant derivative $iD$. 

With this in mind, let us consider $I^{\lambda \mu}$ in
Eq.~(\ref{sloop}). When we first replace  $k$ by the covariant
derivative $\Pi = iD$, and   expand $I^{\lambda \mu}$ in powers of
$\Pi$, we get 
\begin{eqnarray}
I^{\lambda \mu} &=& ieQ_c \int \frac{d^4 l}{(2\pi)^4}
\sum_{m,n=0}^{\infty} (-1)^{m+n+1}  
\frac{1}{\Bigl( (l-q)^2 -m_c^2 \Bigr)^{m+1}}
\frac{1}{\Bigl( l^2 - m_c^2 \Bigr)^{n+1}} \nonumber \\
&\times& \mathrm{tr} \Bigl[ (\FMslash{l} - \FMslash{q} +\FMslash{\Pi}
+m_c ) \Bigl(2\Pi\cdot (l-q) + \Pi^2 -\frac{g}{2} G_{\alpha
\beta}\sigma^{\alpha \beta} \Bigr)^m \nonumber \\
&\times& \gamma^{\lambda} (\FMslash{l}
+\FMslash{\Pi} +m_c) \Bigl( 2\Pi \cdot l +\Pi^2 -\frac{g}{2} G_{\rho
\tau} \sigma^{\rho \tau} \Bigr)^n \gamma^{\mu} (1-\gamma_5) \Bigr],
\label{imunu}
\end{eqnarray}
where $\sigma^{\alpha \beta}= i[\gamma^{\alpha}, \gamma^{\beta}]/2$. 
Here the appearance of $G_{\alpha \beta}$ comes from the $\FMslash{\Pi}
\FMslash{\Pi}$ term when we expand the denominator in
Eq.~(\ref{sloop}) with the identity
\begin{equation}
\FMslash{\Pi} \FMslash{\Pi} = \Pi^2 -\frac{g}{2} G_{\alpha \beta}
\sigma^{\alpha \beta}.
\label{pisquare}
\end{equation}

From Eq.~(\ref{imunu}) we can systematically obtain high-dimensional
operators by a series expansion with respect to $m$ and $n$. Here I
replace the residual momentum by the covariant derivative first and
treat the order of the products carefully as in
Eq.~(\ref{pisquare}). Because of this procedure, there appears
$G_{\alpha \beta}$ in Eq.~(\ref{imunu}).
In the literature \cite{bnon}, there is another way of calculating the
integral. First we expand the diagram in 
powers of the residual momentum $k$, regarding it as a number, i.e.,
$\FMslash{\Pi} \FMslash{\Pi} = \Pi^2$. Then
there appears no $G_{\alpha \beta}$ in the expansion. In order to
include the effect from the gluon field strength tensor $G_{\alpha
\beta}$, we attach an external gluon and extract the term proportional
to $G_{\alpha \beta}$. It is called a ``one-gluon matrix
element''. However it turns out that these two approaches are
equivalent.

It is complicated to obtain a simple and general form for the series
expansion in Eq.~(\ref{imunu}). However we can calculate the first few
high-dimensional operators explicitly in order to see the structure of
them. The zeroth-order term ($m=n=0$) does not contribute to the
high-dimensional operators suppressed by $m_c$.  Also there
is no term linear in  $\Pi$. At second order in $\Pi$, there are no
symmetric terms quadratic in $\Pi$ such as  $\Pi^{\alpha} \Pi^{\beta}
+ \Pi^{\beta} \Pi^{\alpha}$ or $\Pi^2$. The only surviving terms up to
quadratic terms in $\Pi$ are those with $G_{\alpha \beta}$. These
terms come from the series expansion in Eq.~(\ref{imunu}) with $m+n=1$
and 2 respectively. The result is written as
\begin{equation}
\frac{egQ_c}{24\pi^2 m_c^2} \epsilon^{\lambda \alpha \mu \sigma}
q_{\sigma} q^{\beta} G_{\alpha \beta}.
\label{leading}
\end{equation}

Note that all the quadratic contributions which do not
involve $G_{\alpha \beta}$ explicitly in Eq.~(\ref{imunu}) cancel out,
leaving only the one-gluon matrix element. 
The reason why the term with $G_{\alpha \beta}$ alone survives is
clear when we look at the diagram in Fig.~\ref{otwo}. If a single
gluon is attached to the loop, the loop has the 
structure of a vector-vector-axial triangle and only the term with
the Levi-Civita $\epsilon$ tensor and $G_{\alpha \beta}$
survives. The effective operator from Eq.~(\ref{leading}) can be
derived in a straightforward way, and is
given in Eq.~(\ref{new}). Therefore the leading term
of order $1/m_c^2$ coincides with the result by
Voloshin \cite{voloshin}, which is obtained by attaching a soft gluon
to the $c$ quark loop.

Now let us consider higher-order corrections to the leading term given
in Eq.~(\ref{leading}). The operators of $O(1/m_c^4)$ are
obtained from the terms with $m+n=2$ and 3 respectively in
Eq.~(\ref{imunu}). The result is written as 
\begin{equation}
\frac{egQ_c}{180\pi^2 m_c^4} \epsilon^{\lambda \alpha \mu \sigma}
 q_{\sigma} q^{\beta} [q\cdot \Pi , G_{\alpha\beta}].
\label{quartic}
\end{equation}
Recall that the covariant derivative operator $\Pi$ is applied to the
$b$ field. Then the commutator in Eq.~(\ref{quartic}) can be written
as 
\begin{equation}
[q\cdot \Pi , G_{\alpha\beta}] b = (q\cdot \Pi  G_{\alpha\beta}) b,
\label{iden}
\end{equation}
where the covariant derivative $\Pi$ in the right-hand side only
applies to $G_{\alpha\beta}$. Of course, it should be understood that
this holds only for the derivative in the covariant derivative
$\Pi$. Using the identity Eq.~(\ref{iden}), Eq.~(\ref{quartic}) can be
written as 
\begin{equation}
\frac{egQ_c}{180\pi^2 m_c^4} \epsilon^{\lambda \alpha \mu \sigma}
 q_{\sigma} q^{\beta} \bigl( q\cdot \Pi  G_{\alpha\beta}\bigr),
\end{equation}
where $\Pi$ applies only to the gluon field strength tensor.
This is where the HQEFT result is different from the result by
attaching soft gluons to the charm quark loop. To be more
quantitative, the HQEFT result is half 
the higher-dimensional contribution derived from the one-gluon matrix
element and its derivatives, which makes the contribution of
higher-dimensional operators of $O(1/m_c^4)$ smaller than estimated
before \cite{ligeti,grant}.

There are also other operators which cannot be obtained by
attaching external gluons. These operators can be obtained in the
HQEFT scheme. For example, there are higher-dimensional
operators of order $\Pi^3/m_c^2$, though their contributions  are
suppressed by $\Lambda/m_b$  compared to the operators of order
$\Pi^2/m_c^2$.  These terms arise from $m+n=1, 2, 3$
respectively. After a long but straightforward calculation, the result
is given by 
\begin{eqnarray}
&&\frac{eQ_c}{16\pi^2 m_c^2} \Bigl[ \frac{8}{3} i \epsilon^{\alpha
 \lambda \mu \rho} q_{\rho} \Pi_{\alpha} 
\Pi^2 -\frac{g}{6} \epsilon^{\alpha\beta\lambda\mu} [ q\cdot \Pi ,
G_{\alpha \beta}]   \nonumber \\
&-&\frac{g}{3} [\Pi_{\delta}, G_{\alpha \beta}] \Bigl( 6g^{\alpha
\delta} \epsilon^{\beta \lambda\mu\rho} q_{\rho}-4g^{\alpha \mu}
\epsilon^{\beta \delta \lambda \rho} q_{\rho} +
\epsilon^{\beta \delta \lambda \mu} q^{\alpha} \nonumber \\
&+&g^{\delta \lambda} \epsilon^{\alpha \beta \mu \rho} q_{\rho}
-2g^{\delta \mu} \epsilon^{\alpha \beta \lambda \rho} q_{\rho}
+2g^{\mu \lambda}
\epsilon^{\alpha \beta \delta \rho} q_{\rho} -2q^{\mu}
\epsilon^{\alpha \beta \delta \lambda} \Bigr) \Bigr].
\label{picube}
\end{eqnarray}

In evaluating the contribution of higher-dimensional
operators of order $O(\Pi^3)$, Eqs.~(\ref{quartic}) and
(\ref{picube}), to the decay rate, there is one more caveat that we
have to include the contribution of both the operators $O_1$ and
$O_2$. For the leading term, given in Eq.~(\ref{leading}), with one
gluon field strength tensor, only the operator $O_2$ contributes
since it has the correct color structure. However, as mentioned 
before, when we include the emission of two or more gluons, 
the contribution from the operator $O_1$ should also be considered. We
have to separate these higher-dimensional operators into the
color-octet part for $O_2$ and the color-singlet part for
$O_1$. But since these contributions are numerically insignificant,
it will not be pursued here any more.

Implementing the HQEFT idea to the charm quark which is immersed in
the same light degrees of freedom of the decaying $b$ quark, we can
systematically obtain a complete set of high-dimensional operators
suppressed by powers of $m_c$. The effect of attaching external
gluons gives the correct leading term, but it does not include all
the higher-dimensional operators. In other words, attaching external 
gluons corresponds to the expansion in powers of the gluon field
strength tensor $G_{\alpha \beta}$ and its derivatives. The resultant
operators for $n$ attached gluons are symbolically of the form
$G_{\alpha \beta}^n$ with some derivatives. In order to obtain a
complete set of high-dimensional operators, the HQEFT idea can be
applied to the decay $B\rightarrow X_s \gamma$. 

The numerical estimate of the higher-dimensional operators are
negligible compared to the leading operator in Eq.~(\ref{new}), which 
already contributes about 3\% to the decay rate. 
As pointed out in Refs.~\cite{ligeti,grant,rey}, it is
not because the expansion parameter $m_b \Lambda_{\mathrm{QCD}}
/m_c^2$ is small, but because the coefficients are small. 
Even though higher-dimensional operators are numerically
insignificant, the idea of the HQEFT offers a systematic way of
obtaining a complete set of high-dimensional operators.

Conceptually there is an advantage in considering the nonperturbative 
contributions using the HQEFT idea. The effect of the operators
suppressed by powers of $m_c$ is  nonperturbative. The reason
why we could systematically classify high-dimensional operators is
because the charm quark is heavy compared to the QCD scale
$\Lambda_{\mathrm{QCD}}$. If the charm quark is not heavy, it is
impossible to analyze the $1/m_c$ effects using the OPE, and this
effect belongs to a purely nonperturbative regime. And it is not
surprising to have nonperturbative contributions governed by the
scale $m_c$ as well as $m_b$ in $B$ decays since these effects arise 
simply because the $c$ quark is also heavy. In the same line of
reasoning, it is impossible to analyze the contribution of the up
quark loop from a similar process in inverse powers of the up quark
mass using the OPE since this contribution is genuinely
nonperturbative.  

\begin{ack}
The author is grateful to Adam Falk, Seyong Kim, Pyungwon Ko, Soo-Jong
Rey for helpful discussion. This work is supported in part by the
Ministry of Education grants BSRI 97-2408,
the Korea Science and Engineering Foundation (KOSEF) through the SRC
program of SNU-CTP, the Distinguished Scholar Exchange Program of
Korea Research Foundation and the German-Korean scientific exchange
program DFG-446-KOR-113/72/0.
\end{ack}


\begin{thebibliography}{99}
\bibitem{cleo} CLEO Collab., M. S. Alam et al., Phys.
Rev. Lett. {\bf 74} (1995) 2885.
\bibitem{kagan} A. L. Kagan, M. Neubert, CERN-TH/98-1, UCHEP-98/7
[hep-ph/9803368]; CERN-TH/98-99 [hep-ph/9805303].
\bibitem{matching} K. Adel and A. P. Yao, Phys. Rev. {\bf D49} (1994)
4945; C. Greub and T. Hurth, Phys. Rev. {\bf D56} (1997) 2934.
\bibitem{mat} A. Ali and C. Greub, Phys. Lett. {\bf B361} (1995) 146;
C. Greub, T. Hurth and D. Wyler, Phys. Lett. {\bf B380} (1996) 385;
N. Pott, Phys. Rev. {\bf D54} (1996) 938.
\bibitem{anom} K. Chetyrkin, M. Misiak and M. M\"{u}nz,
Phys. Lett. {\bf B400} (1997) 206. 
\bibitem{chay} J. Chay, H. Georgi and B. Grinstein, Phys. Lett. {\bf
B247} (1990) 399.
\bibitem{bnon} A. V. Manohar and M. B. Wise, Phys. Rev. {\bf D49}
(1994) 1310; A.F. Falk, M. Luke and M.J. Savage, Phys. Rev. {\bf
D49} (1994) 3367.
\bibitem{grin} B. Grinstein, R. Springer and M. B. Wise,
Nucl. Phys. {\bf B339} (1990) 269.
\bibitem{voloshin}  M. B. Voloshin, Phys. Lett. {\bf B397} (1997) 275.
\bibitem{ligeti} Z. Ligeti, L. Randall and M. B. Wise,
Phys. Lett. {\bf B402}, 178 (1997). 
\bibitem{grant} A. K. Grant, A. G. Morgan, S. Nussinov and
R. D. Peccei, Phys. Rev. {\bf D56} (1997) 3151.
\bibitem{wyler} A. Khodjamirian, R. R\"{u}ckl, G. Stoll and D. Wyler,
Phys. Lett. {\bf B402} (1997) 167. 
\bibitem{rey} G. Buchalla, G. Isidori and S.-J. Rey, Nucl. Phys. {\bf
B511} (1998) 594.
\end{thebibliography}
\end{document}